\documentclass[11pt,a4paper]{article}
\pdfoutput=1

\usepackage{jheppub}
\usepackage[lofdepth,lotdepth]{subfig} 
\usepackage[pass]{geometry}
\usepackage{adjustbox}

\usepackage{import}

\usepackage{bigints}

\usepackage{array}
\newcolumntype{C}[1]{>{\centering\arraybackslash}m{#1}}

\title{On the Boundaries of the m=2 Amplituhedron}
\author[1]{Tomasz \L ukowski}\emailAdd{t.lukowski@herts.ac.uk}
\affiliation[1]{School of Physics, Astronomy and Mathematics, \\
 University of Hertfordshire, \\
 Hatfield, Hertfordshire, AL10 9AB, UK}
\abstract{Amplituhedra $\mathcal{A}_{n,k}^{(m)}$ are geometric objects of great interest in modern mathematics and physics:  for mathematicians they are combinatorially rich generalizations of polygons and polytopes, based on the notion of positivity; for physicists, the amplituhedron $\mathcal{A}^{(4)}_{n,k}$ encodes the scattering amplitudes of the planar $\mathcal{N}=4$ super Yang-Mills theory. In this paper we study the structure of boundaries for the amplituhedron $\mathcal{A}_{n,k}^{(2)}$. We classify all boundaries of all dimensions and provide their graphical enumeration. We find that the boundary poset for the amplituhedron is Eulerian and show that the Euler characteristic of the amplituhedron equals one. This provides an initial step towards proving that the amplituhedron for $m=2$ is homeomorphic to a closed ball.}

\begin{document}
\begin{flushright}
{\small }
\end{flushright}
\maketitle

\section{Introduction}
It is an elementary fact about convex polygons that they are homeomorphic to a two-dimensional ball: a unit disk. This is strongly reflected in the combinatorial structure of their boundaries and allows to prove general statements about them, for example that their Euler characteristic equals one. For polygons this fact is easy to show by direct counting: each $n$-gon has exactly $n$ edges and exactly $n$ vertices which leads us to the Euler characteristic
\begin{equation}
  \chi=  1-n+n=1\,,
\end{equation}
as expected. However, such direct enumeration is more involved for higher-dimensional convex polytopes, which nonetheless are homeomorphic to $d$-dimensional spheres, and therefore their Euler characteristic has to be equal to  1. In recent years it was shown that amplituhedra defined in \cite{Arkani-Hamed:2013jha} share many features with polygons and polytopes. Amplituhedra $\mathcal{A}_{n,k}^{(m)}$ are generalizations of polytopes into a Grassmannian space, relying on the notion of positivity, and they exhibit a rich combinatorial structure. They can be defined as the image of the positive Grassmannian through a positive linear map, and similar to polytopes they are bounded regions equipped with cell decompositions. They are also of great interest in high-energy physics since they provide a geometric description for tree-level scattering amplitudes in planar $\mathcal{N}=4$ super Yang-Mills (SYM). There, positivity properly encodes the factorization properties of amplitudes, and replaces  unitarity and locality as fundamental notions, providing a surprising new framework for quantum theories.

Despite displaying a complicated combinatorial structure, amplituhedra are rather simple topologically.  
In particular, it was proven in \cite{Galashin:2017onl} that the $m=1$ amplituhedron $\mathcal{A}_{n,k}^{(1)}$ is homeomorphic to a $k$-dimensional ball. Our conjecture is that this is also true  for $m=2$, and in this paper we take a first step towards confirming this statement. We start by classifying all boundaries of the amplituhedron $\mathcal{A}_{n,k}^{(2)}$ and study the structure of their partially ordered set ({\em poset}) of boundaries. We check that this poset is Eulerian and find that the Euler characteristic of $\mathcal{A}_{n,k}^{(2)}$ equals one. We also provide a useful diagrammatic enumeration for all boundaries, which enables us to study combinatorial properties of $\mathcal{A}_{n,k}^{(2)}$ for any $n$ and $k$.  

\section{Amplituhedra and Their Boundaries}
\subsection{Positive Grassmannian}
Our approach will strongly rely on the known classification of boundaries of the positive Grassmannians \cite{Postnikov2006,ArkaniHamed:2012nw}, which states that  each boundary of positive Grassmannian $G_+(k,n)$ can be parametrized by a permutation of $n$ elements. Let us denote by $\Sigma_{n,k}$ the set of all positroid cells for $G_{+}(k,n)$, i.e.~the boundaries of the positive Grassmannian $G_{+}(k,n)$, of all possible dimensions. We also include the top cell of $G_{+}(k,n)$ in $\Sigma_{n,k}$.  There is a natural partial order $\prec_C$ on $\Sigma_{n,k}$ given by:
\begin{equation}
\sigma_1\prec_C\sigma_2 \text{ iff } \sigma_1 \text{ is a boundary of } \sigma_2.
\end{equation}
We can extend this order transitively to obtain a partially ordered set of positive Grassmannian boundaries $(\Sigma_{n,k},\prec_C)$. This poset is graded by the cell dimension.
For each positroid cell $\sigma$ we define two sets:
\begin{itemize}
\item $\partial_C \sigma$ is the set of all boundaries of $\sigma$ (of all dimensions): $\partial_C\sigma=\{\sigma'\in \Sigma_{n,k}:\sigma'\prec_C\sigma\}$
\item $\partial^{-1}_C \sigma$ is the set of all {\em inverse boundaries} of $\sigma$, i.e.~the set of all positroid cells $\sigma'$ for which $\sigma\in\partial_C\sigma'$: $\partial^{-1}_C\sigma=\{\sigma'\in \Sigma_{n,k}:\sigma\prec_C\sigma'\}$.
\end{itemize}

\subsection{Amplituhedron}
The amplituhedron $\mathcal{A}_{n,k}^{(m)}$ is defined as the image of the positive Grassmannian $G_+(k,n)$ through the function
\begin{equation}
    \Phi_Z:G_{+}(k,n)\to G(k,k+m)\,,
\end{equation}
induced by  a $(k+m)\times n$ matrix $Z$ with positive maximal minors, defined by
\begin{equation}
  Y_\alpha^A= \Phi_Z(C)=c_{\alpha i}Z_i^A\,,\qquad\mbox{for } C=\{c_{\alpha i}\}\in G_{+}(k,n)\,.
\end{equation}
In this paper we focus on the case $m=2$. For $k=1$ the image is the familiar planar, convex $n$-gon embedded in the projective space $\mathbb{P}^2$. When $k>1$, it is a more complicated bounded region inside the Grassmannian $G(k,k+2)$ with all co-dimension one boundaries given by
\begin{equation}
\langle Yii+1\rangle=0\,,
\end{equation}
where $\langle Yij\rangle=\epsilon_{A_1\ldots A_k BC}Y_1^{A_1}\ldots Y_k^{A_k}Z_i^BZ_j^C $.
The structure of all lower-dimensional boundaries of $\mathcal{A}_{n,k}^{(2)}$ has not been studied before and this paper provides their full classification.

\subsection{Amplituhedron Dimension}
As we already mentioned, the full stratification of the positive Grassmannian $G_{+}(k,n)$ is known. For a given positroid cell $\sigma$ we can define its {\em Grassmannian dimension}, which we will denote by $\dim_{C}(\sigma)$. It is important to notice that the dimension of the image of a positroid cell through the map $\Phi_Z$ can differ from the dimension of the cell itself. It can be already observed for $n$-gons, i.e.~the $k=1$ case: although the dimension of the top cell of the positive Grassmannian $G_{+}(1,n)$ equals $n-1$, the image through the map $\Phi_Z$ is always two-dimensional. We therefore define the {\it amplituhedron dimension} $\dim_A(\sigma)$ of a positroid cell $\sigma$  as the dimension of its image:
\begin{equation}
\dim_A(\sigma)=\dim(\Phi_Z(\sigma))\,.
\end{equation}
Since it is always true that $\dim_C(\sigma)\geq \dim_A(\sigma)$, we can distinguish two cases:
\begin{itemize}
\item $\dim_C(\sigma)=\dim_A(\sigma)$: this will correspond to a simplicial-like image,
\item $\dim_C(\sigma)>\dim_A(\sigma)$: the image will be polytopal-like.
\end{itemize}
In particular, for a positroid cell $\sigma$ for which $\dim_C(\sigma)>\dim_A(\sigma)$, we can find a collection of cells in $\partial_C\sigma$ with the same amplituhedron dimension as $\sigma$. Moreover, in all cases we studied, there exists a (non-unique) subset $\{\sigma_1,\ldots,\sigma_r\}\in\partial_C\sigma$ such that the images $\{\Phi_Z(\sigma_1),\ldots,\Phi_Z(\sigma_r)\}$ triangulate the image $\Phi_Z(\sigma)$. In the generic case when $n>m+k$, the image of the top cell of $G_{+}(k,n)$, i.e.~the amplituhedron $\mathcal{A}_{n,k}^{(2)}$ itself, is polytopal-like and we are often interested in finding a collection of $(k\cdot m)$-dimensional cells which triangulate it. We want to emphasize here that a similar behaviour is also true for many boundaries of the amplituhedron.

\subsection{Amplituhedron Boundaries}
In this paper we are interested in studying the boundary structure of the amplituhedron $\mathcal{A}_{n,k}^{(2)}$. Let us denote by $\mathcal{B}_{n,k}$ the set of all boundaries of $\mathcal{A}_{n,k}^{(2)}$. As we discussed in the previous section, we can encounter two types of boundaries: simplicial-like and polytopal-like. In the former case we can label any simplicial-like boundary $B_\sigma\in\mathcal{B}_{n,k}$ by the corresponding positroid cell for which $B_\sigma=\Phi_Z(\sigma)$, and therefore by the permutation associated to $\sigma$.  For a $d$-dimensional polytopal-like boundary $\tilde B\in \mathcal{B}_{n,k}$ the situation is more complicated since there will be many positroid cells, with amplituhedron dimension $d$, which will be mapped to (a subset of) $\tilde B$. In order to find a unique label for each boundary, we will characterize $\tilde B$ by the positroid cell which is mapped to $\tilde B$ with the highest Grassmannian dimension. In particular, it implies that the interior of the amplituhedron $\mathcal{A}_{n,k}^{(2)}$ is labelled by the permutation of the top cell of $G_+(k,n)$. Importantly, the dimension of a boundary can be much smaller than the Grassmannian dimension of the associated cell. 

As we already mentioned, the co-dimension one boundaries of the amplituhedron are well understood and they take the form $ \langle Y i i+1\rangle =0$. To find lower-dimensional boundaries we will proceed recursively: assume that we have found all amplituhedron boundaries of dimension larger than $d$.  Let us study all positroid cells $\sigma\in \Sigma_{n,k}$ with amplituhedron dimension $\dim_A \sigma=d$. For a given cell $\sigma$, there are two options: 
\begin{itemize}
\item either the amplituhedron dimension for all inverse boundaries of $\sigma$ are higher than the amplituhedron dimension of $\sigma$: $\forall_{\sigma'\in\partial^{-1}\sigma}: \dim_A\sigma'>\dim_A\sigma$;
\item or we can find a cell among the inverse boundaries of $\sigma$ which has a higher Grassmannian dimension but the same amplituhedron dimension as $\sigma$: $\exists_{\sigma'\in\partial^{-1}\sigma}:\dim_A\sigma'=\dim_A\sigma \mbox{ and } \dim_C\sigma\prec_C \dim_C\sigma'$.
\end{itemize}
We only keep the former cells, since the latter are necessarily elements of a triangulation of a boundary of the amplituhedron. After doing that, there is still a possibility that some of the remaining cell images are spurious boundaries, which arise as spurious faces in triangulations of polytopal-like boundaries.
Spurious boundaries can be identified (and removed) because they belong to a single $(d+1)$-dimensional amplituhedron boundary, while external boundaries belong to at least two such boundaries. This procedure allows us to find all external boundaries of dimension $d$. We can follow this procedure recursively, starting from the known co-dimension one boundaries, and work our way down to zero-dimensional boundaries: points. 

\subsection{Graphical Notation}
We have followed the procedure described above and studied the set of all amplituhedron boundaries $\mathcal{B}_{n,k}$ for a wide range of $n$ and $k$. It led us to a graphical notation for all boundaries of amplituhedron $\mathcal{A}_{n,k}^{(2)}$ which we describe in the remaining part of this section. 

\begin{figure}[t]
\begin{center}
\begin{tabular}{ccc}
\raisebox{1.2cm}{$P^{(n)}_{top}=$}\includegraphics[scale=0.2]{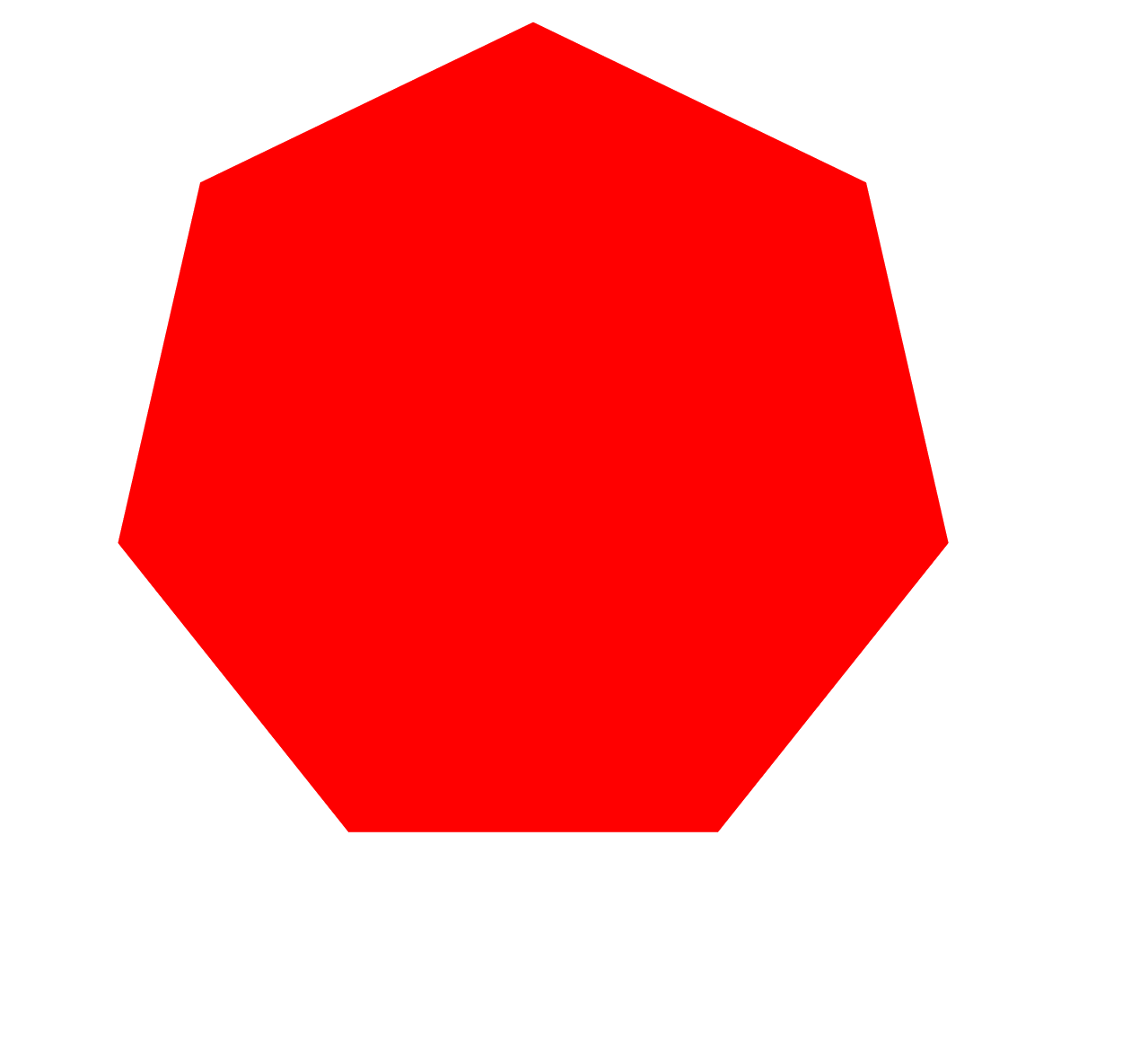}\hspace{1cm}\phantom{.}&\raisebox{1.2cm}{$P^{(n)}_{i,i+1}=$}\includegraphics[scale=0.2]{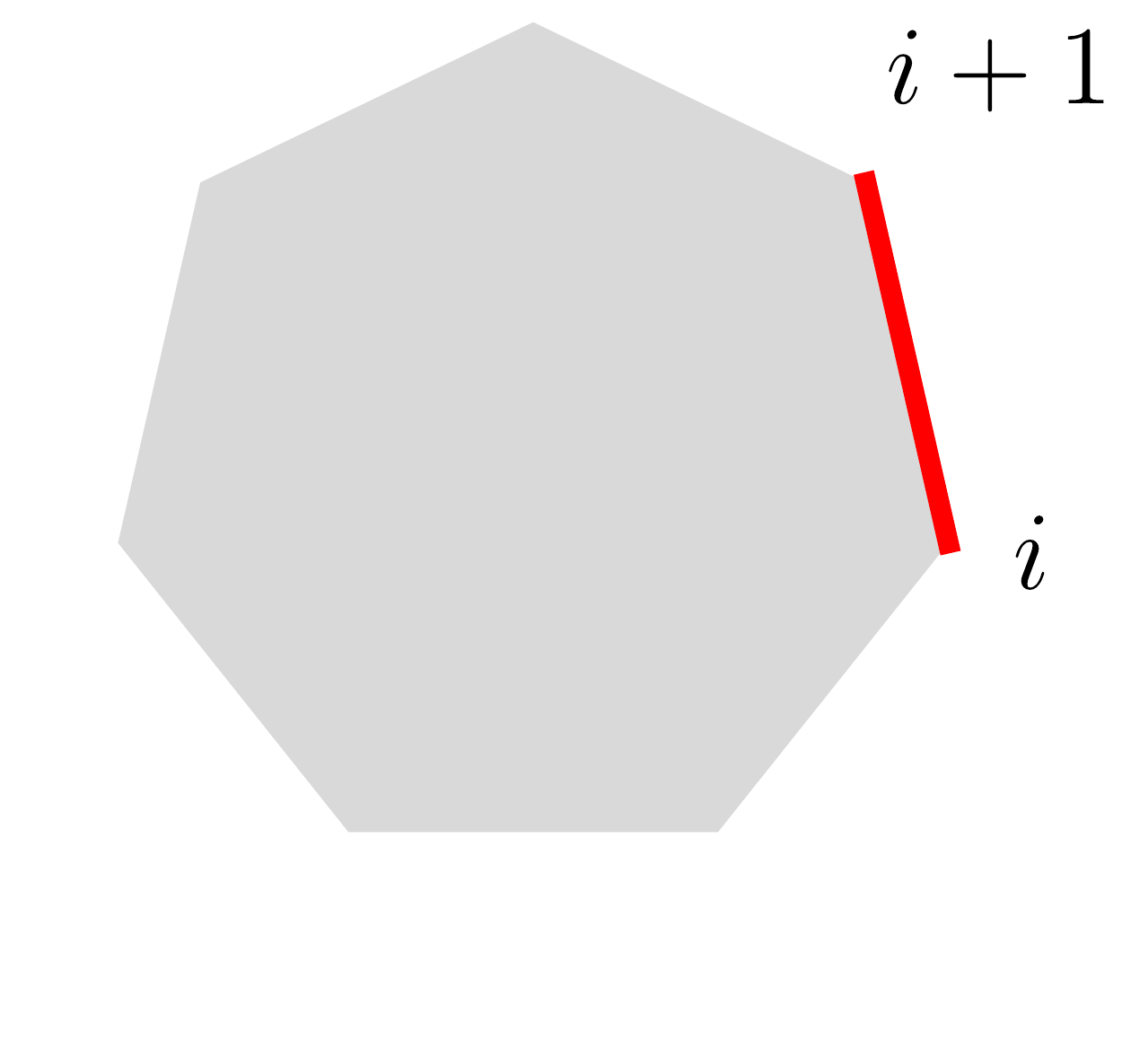}\hspace{1cm}\phantom{.}&\raisebox{1.2cm}{$P^{(n)}_{i}=$}\includegraphics[scale=0.2]{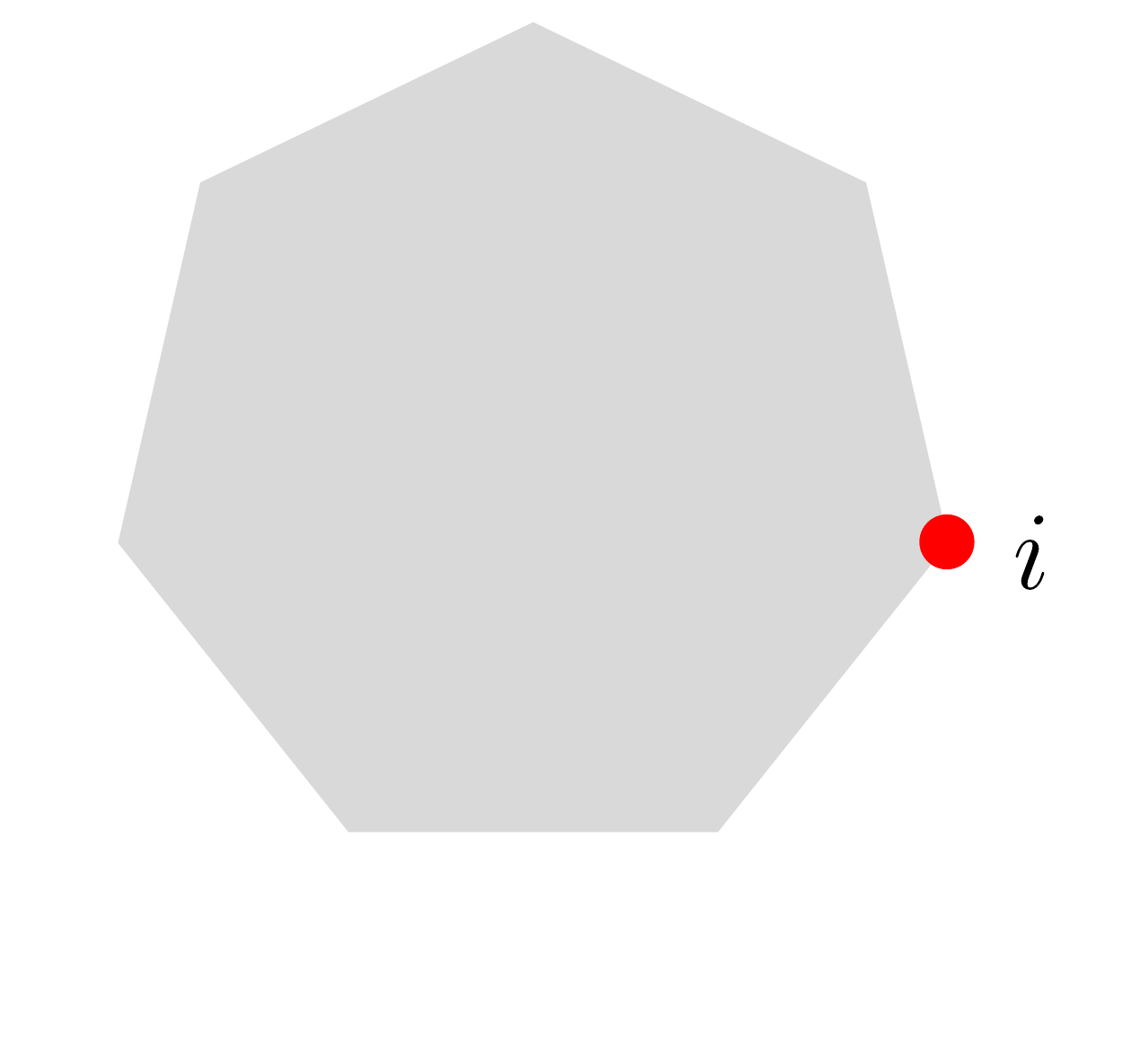}\\
a)\hspace{1cm}\phantom{.}&b)\hspace{1cm}\phantom{.}&c)
\end{tabular}
\end{center}
\caption{Labels for the amplituhedron $\mathcal{A}_{n,1}^{(2)}$: a) two-dimensional bulk, b) one-dimensional boundaries (lines $(i,i+1)$), c) zero-dimensional boundaries (points $i$)}
\label{Fig:k1}
\end{figure}

First, it is easy to enumerate all boundaries of polygons, i.e. the case $k=1$. There are exactly $n$ co-dimension one boundaries: edges, and exactly $n$ co-dimension two boundaries: points. We will label them as in Fig.~\ref{Fig:k1}, where we introduced a label for the interior of the $n$-gon: $P^{(n)}_{top}$, and its pictorial label as a solid $n$-gon; the labels for edges: $P^{(n)}_{i,i+1}$, which are represented pictorially but highlighting the edge $(i,i+1)$; and for vertices: $P^{(n)}_{i}$, with a vertex $i$ highlighted in its graphical label.
Both notations can be generalized to higher $k$ to enumerate all boundaries in $\mathcal{B}_{n,k}$. We explain the main features of our notation by focusing on $k=2$, which already provides a new and rich structure, at the same time highlighting all types of behaviour we encounter for higher $k$. The amplituhedron $\mathcal{A}_{n,2}^{(2)}$ itself, which is the four-dimensional image of the top cell of $G_+(2,n)$ through $\Phi_Z$, is labelled by two solid $n$-gons:
\begin{center}
\raisebox{1.2cm}{$P^{(n)}_{top}\otimes P^{(n)}_{top}=$}\includegraphics[scale=0.2]{polyfull.pdf}\includegraphics[scale=0.2]{polyfull.pdf}
\end{center}
Co-dimension one boundaries, corresponding to $\langle Yii+1\rangle=0$, are depicted as
\begin{center}
\raisebox{1.2cm}{$P^{(n)}_{top}\otimes P^{(n)}_{i,i+1}=$}\includegraphics[scale=0.2]{polyfull.pdf}\includegraphics[scale=0.2]{polyline.pdf}
\end{center}
There are two types of co-dimension two boundaries:
\begin{center}
\raisebox{1.2cm}{$P^{(n)}_{top}\otimes P^{(n)}_{i}=$}\includegraphics[scale=0.2]{polyfull.pdf}\includegraphics[scale=0.2]{polypoint.pdf}
\end{center}
and
\begin{center}
\raisebox{1.2cm}{$P^{(n)}_{i,i+1}\otimes P^{(n)}_{j,j+1}=$}\includegraphics[scale=0.2]{polyline.pdf}\includegraphics[scale=0.2]{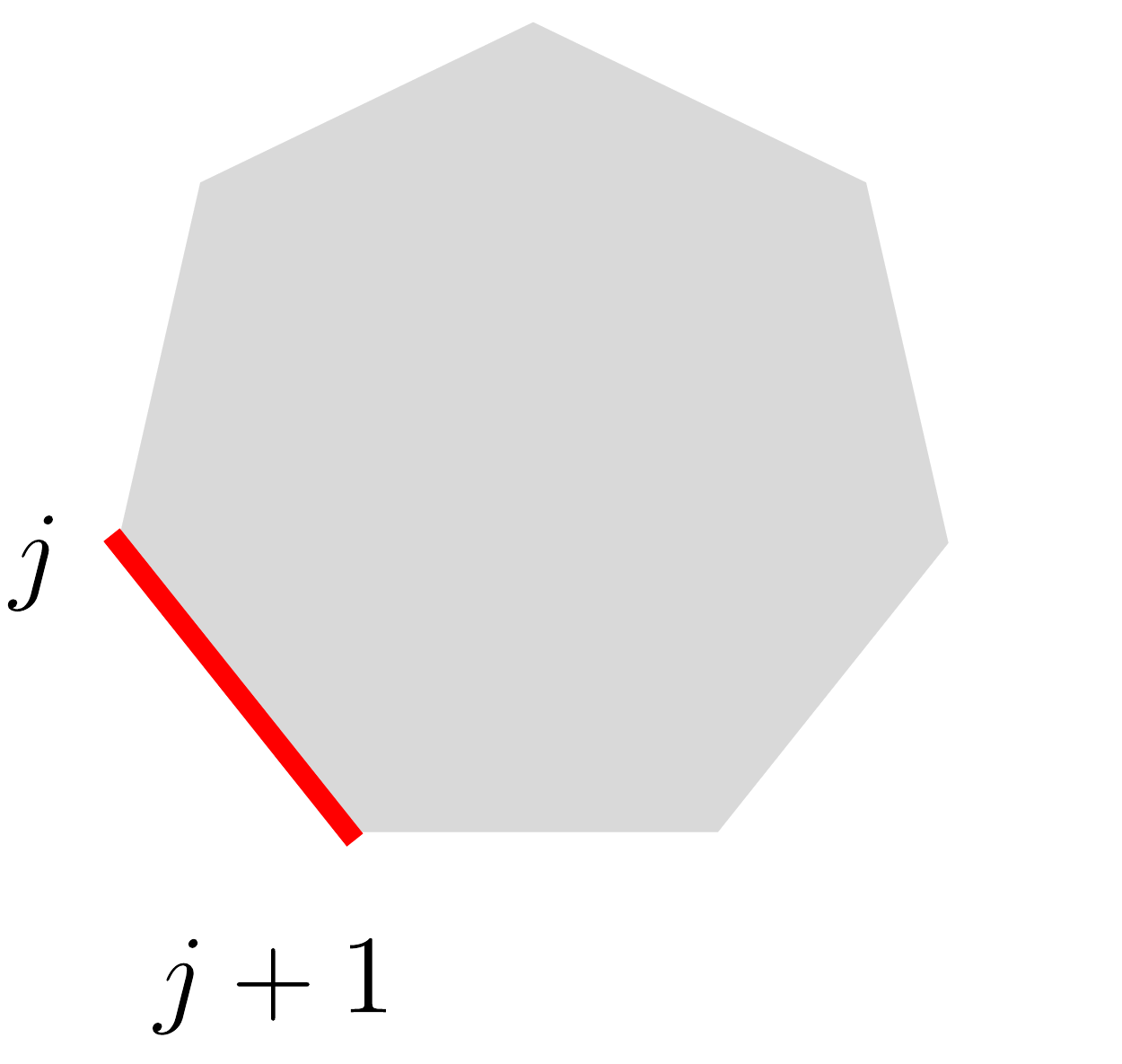}
\end{center}
with $i< j$. There is again only one type of co-dimension three boundaries which in the generic case can be labelled as
\begin{center}
\raisebox{1.2cm}{$P^{(n)}_{i,i+1}\otimes P^{(n)}_{j}=$}\includegraphics[scale=0.2]{polyline.pdf}\includegraphics[scale=0.2]{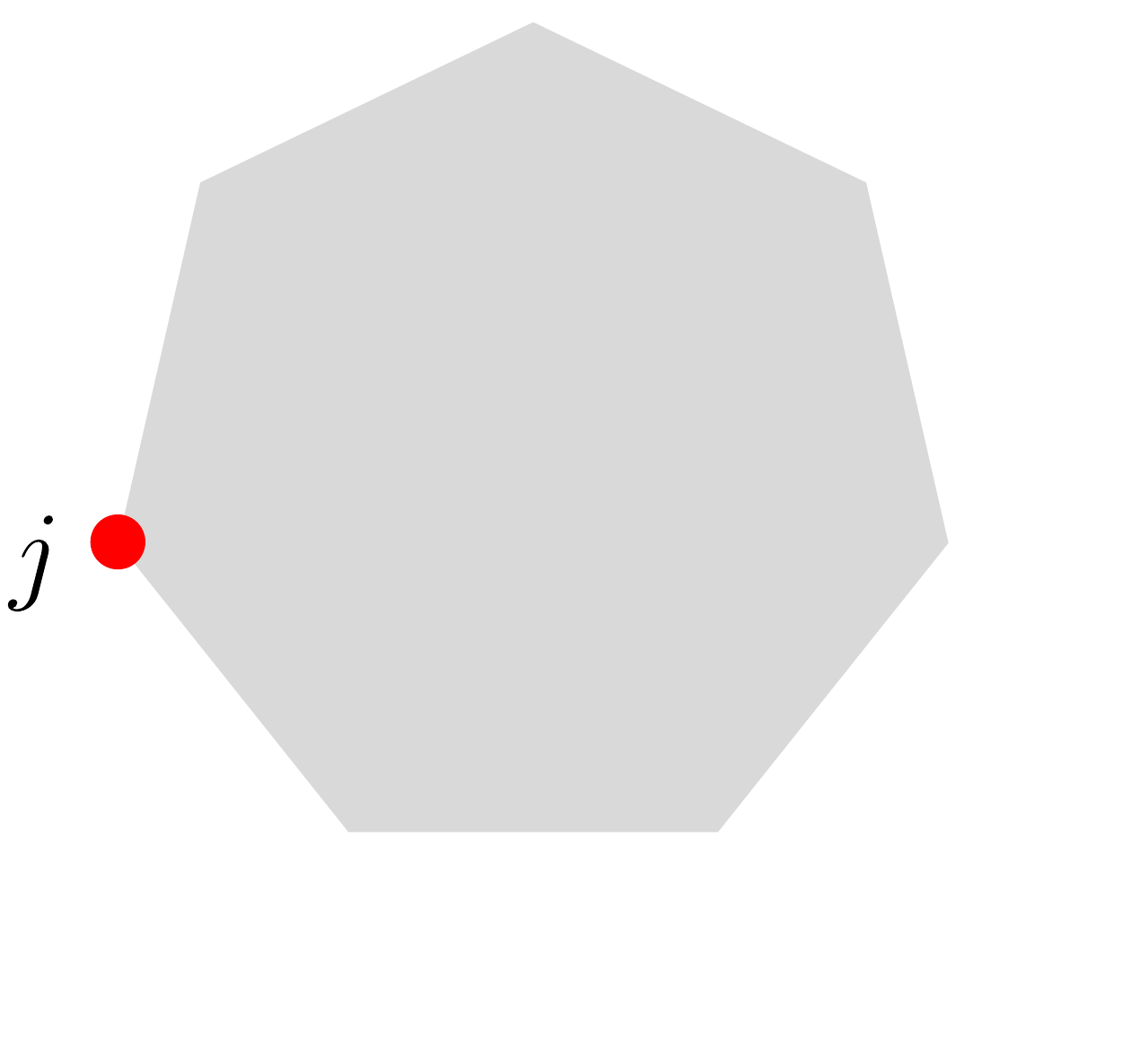}
\end{center}
with $j\neq i,i+1$.
We also find an additional type of labels for non-generic boundaries\footnote{We interpret these labels for non-generic boundaries in the following way: whenever we encounter $P^{(n)}_i$ in our label, we need to remove the point $i$ from our considerations and in all remaining labels we turn the $n$-gon into the $(n-1)$-gon with the point $i$ removed. Then the boundary $(i-1,i+1)$ is one of the boundaries of the remaining $(n-1)$-gon.}:
\begin{center}
\raisebox{1.2cm}{$P^{(n)}_{i-1,i+1}\otimes P^{(n)}_{i}=$}\includegraphics[scale=0.2]{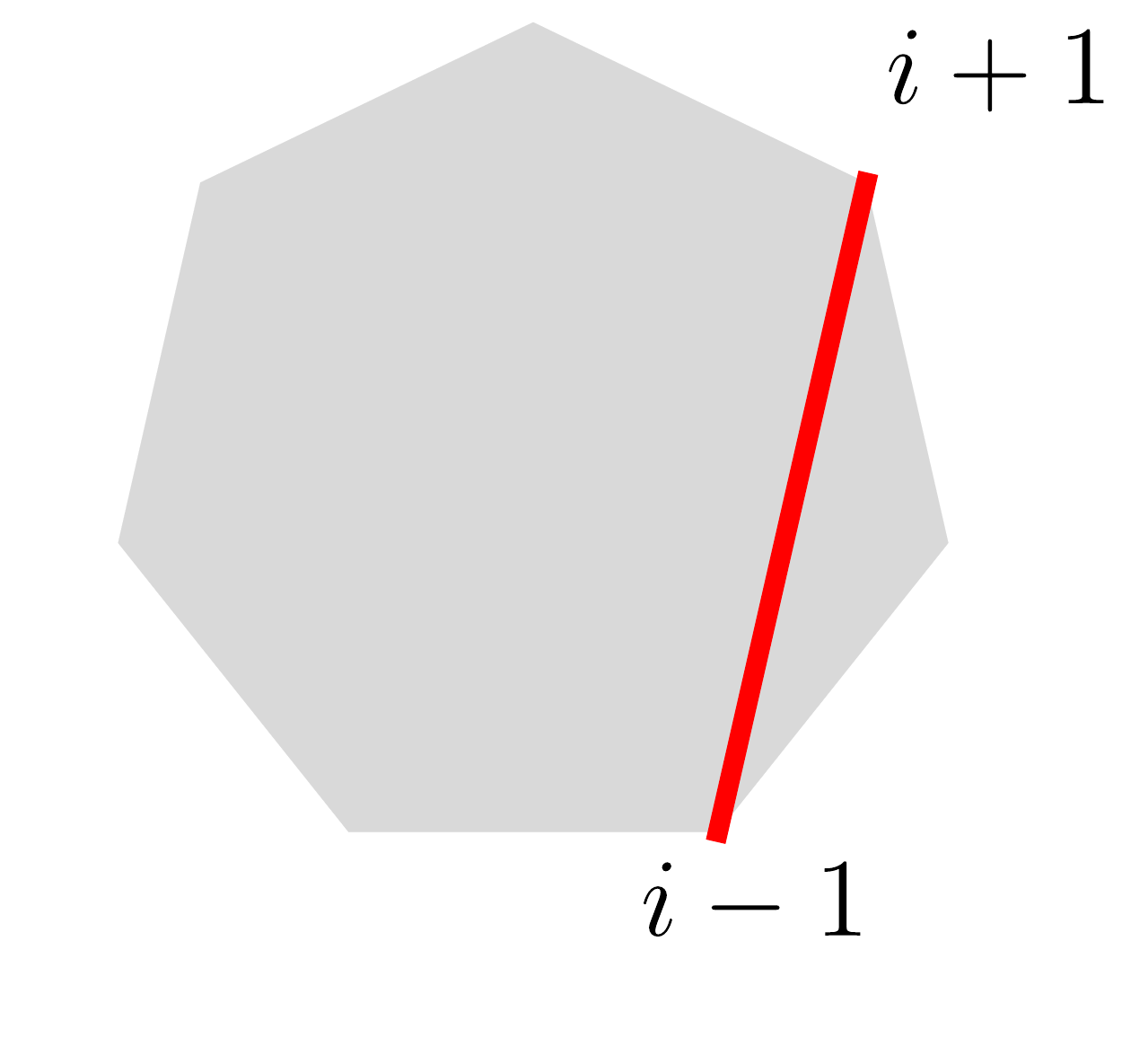}\includegraphics[scale=0.2]{polypoint.pdf}
\end{center}
 Finally, co-dimension four boundaries are just the amplituhedron vertices and they can be depicted as
\begin{center}
\raisebox{1.2cm}{$P^{(n)}_{i}\otimes P^{(n)}_{j}=$}\includegraphics[scale=0.2]{polypoint.pdf}\includegraphics[scale=0.2]{polypointj.pdf}
\end{center}
with $i<j$.

This diagrammatics can be easily generalized to higher $k$ and any boundary of $\mathcal{A}_{n,k}^{(2)}$ can be labelled by $k$ copies of an $n$-gon:
\begin{equation}\label{general_label}
\underbrace{P^{(n)}_{top}\otimes \ldots\otimes P^{(n)}_{top}}_{t-times}\otimes P^{(n)}_{i_1 i_1^+}\otimes\ldots P^{(n)}_{i_l i_l^+}\otimes P^{(n)}_{j_1}\otimes\ldots \otimes  P^{(n)}_{j_p}\,,
\end{equation}
where $t+l+p=k$. The indices are ordered: $i_1<i_2<\ldots<i_l$ and $j_1<j_2<\ldots<j_p$, and we assume $i_a\neq j_b$ for $a=1,\ldots,l$ and $b=1,\ldots,p$. Here, $i_a^+$ is the smallest number in the ordered set $\{i_a+1,\ldots,n,1,\ldots_,i_a\}\setminus\{j_1,\ldots,j_p\}$. To illustrate our diagrammatics, we provide the complete list of all boundaries of the amplituhedron $\mathcal{A}_{5,2}^{(2)}$ in Fig. \ref{Fig:n5k2}. We also include for every boundary the permutation of the corresponding positroid cells of the positive Grassmannian $G_+(2,5)$. The diagrammatic notation we just introduced, apart from providing a full list of boundaries of the amplituhedron $\mathcal{A}_{n,k}^{(2)}$, also encodes many properties of each boundaries, as we explain in Appendix \ref{App:labels}. 

To conclude this section, we introduce a simplified version of our diagrammatics. When working with fixed $k$, and since the marked edges and points present in different $n$-gons do not intersect or overlap, we can simplify our notation by combining all $n$-gons into one, e.g.~for $k=4$ we have:
\begin{center}
\includegraphics[scale=0.2]{polyfull.pdf}\includegraphics[scale=0.2]{polyfull.pdf}\includegraphics[scale=0.2]{polylinej.pdf}\includegraphics[scale=0.2]{polypoint.pdf}\quad\raisebox{1.2cm}{$\longrightarrow$} \quad\includegraphics[scale=0.2]{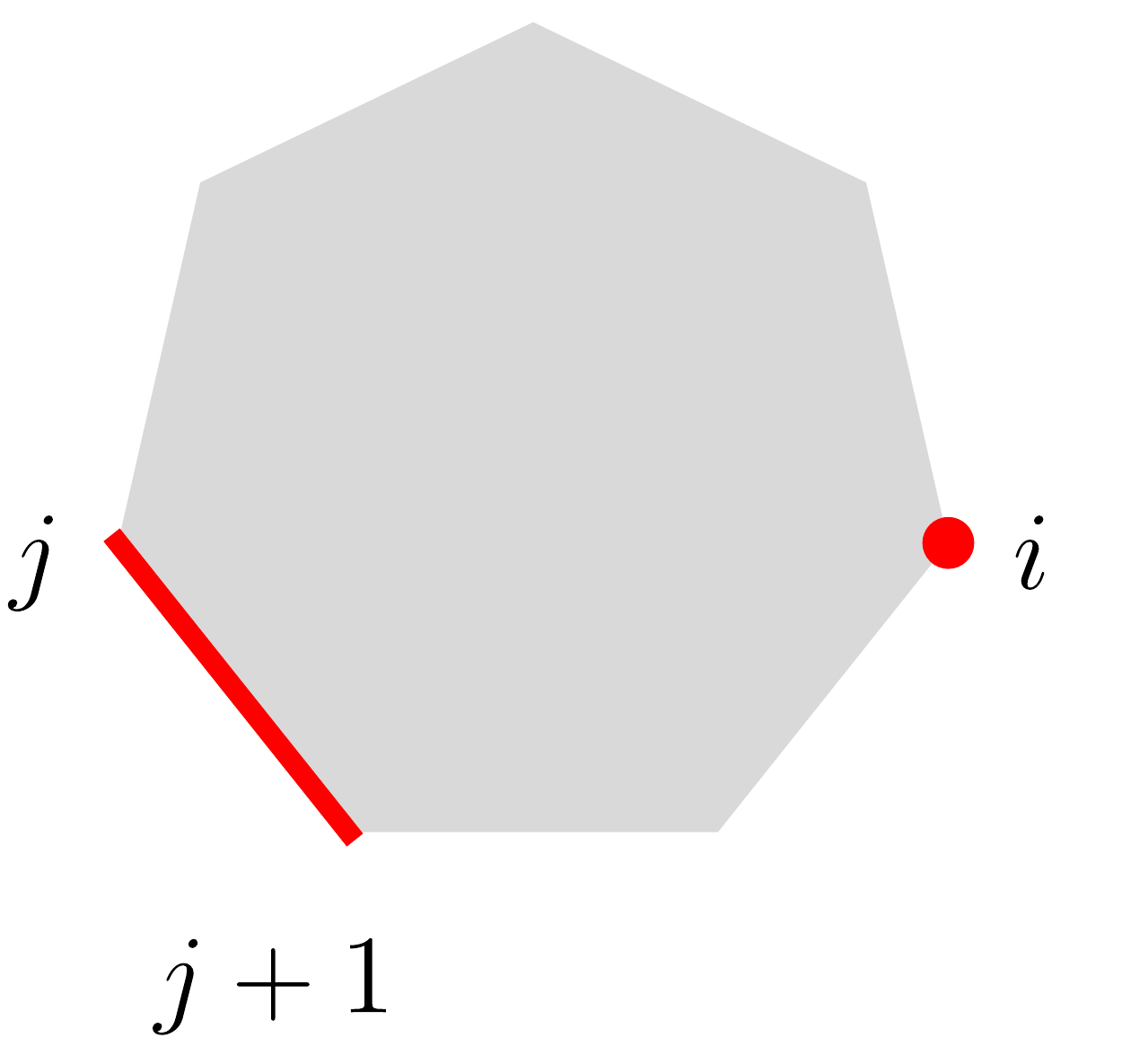}
\end{center}
It is always possible to read off the full label from the simplified one if we know the value of $k$. One needs to separate all line and points and place them on different copies of an $n$-gon, and then add a number of $P^{(n)}_{top}$ such that there are $k$ copies of $n$-gon in total.

\subsection{Amplituhedron Boundary Poset}
As for the positive Grassmannian we can also define a poset of boundaries for the amplituhedron $\mathcal{A}_{n,k}^{(2)}$. First, we define a partial order on the set of all boundaries $\mathcal{B}_{n,k}$ by
\begin{equation}
 B_1\prec_A B_2 \text{ iff } B_1 \text{ is a boundary of } B_2.
\end{equation}
and we extend it transitively. Clearly, if  $B_1\prec_A B_2$ then $\dim_A B_1 < \dim_A B_2$. This poset is graded by the amplituhedron dimension. It allows us to depict it as a layered graph, where each layer corresponds to boundaries with a given dimension. An example of such poset graph can be found in Fig.~\ref{Fig:posetn5k2}, where we study the amplituhedron $\mathcal{A}_{5,2}^{(2)}$ (we use the reduced graphical notation there).

One can check that the amplituhedron boundary poset $(\mathcal{B}_{n,k},\prec_A)$ possesses an interesting property, namely that it is Eulerian. Eulerian posets are important notions in topological combinatorics and are generalizations of boundary posets of polytopes. To define them we need to introduce the notion of a poset interval: for all $B_1\prec_A B_2$  we denote by $[B_1,B_2]$ the {\em interval} from $B_1$ to $B_2$, that is the set $\{B\in \mathcal{B}_{n,k}:B_1\prec_A B\prec_A B_2\}$. A poset is called {\em Eulerian} if every interval of length at least one has the same number of elements of odd rank as of even rank. For us, the rank is defined as the amplituhedron dimension of a given boundary.
It was shown in \cite{Williams2005} that the positive Grassmannian boundary poset $(\Sigma_{n,k},\prec_C)$ is Eulerian. More recently, it has been proven that the amplituhedron $\mathcal{A}_{n,k}^{(1)}$ is homeomorphic to a ball \cite{Galashin:2017onl}, which also implies that its boundary poset is Eulerian. We have checked for various values of $n$ and $k$ that the poset $(\mathcal{B}_{n,k},\prec_A)$ of the amplituhedron $\mathcal{A}_{n,k}^{(2)}$ is Eulerian as well. With the use of our diagrammatic notation, it seems to be within reach to find a rigorous proof of this statement. We leave this problem for future work.

\subsection{Counting Boundaries}
Using the method described in previous sections we can find all boundaries of a given amplituhedron $\mathcal{A}_{n,k}^{(2)}$ and label them using their graphical representatives or the permutations of the corresponding positroid cells. In this section we use this diagrammatics to calculate how many boundaries of a given dimension one can find for a given amplituhedron $\mathcal{A}_{n,k}^{(2)}$. This will allow us to calculate the Euler characteristic $\chi_{n,k}$ for each amplituhedron, which turns out to be equal to one for all $n$ and $k$. We collect our results in the tables below, where we gather all different types of boundaries for $\mathcal{A}_{n,k}^{(2)}$, for $k=1,2,3$. We also include the number of boundaries of a given type and indicated the Grassmannian dimension of all cells corresponding to these boundaries. Starting with $k=1$, we simply obtain the counting for an $n$-gon, as shown in Table \ref{Tab:k1}.
\begin{table}[t]
\begin{center}
\begin{tabular}{|c|c|c|c|}
\hline
$\dim_A$&boundary type&number of boundaries&$\dim_C$\\
\hline
$2$&\adjustbox{raise=-0.25cm}{\includegraphics[scale=0.05]{polyfull.pdf}}&$1$&$n-1$\\
\hline
$1$&\adjustbox{raise=-0.25cm}{\includegraphics[scale=0.05]{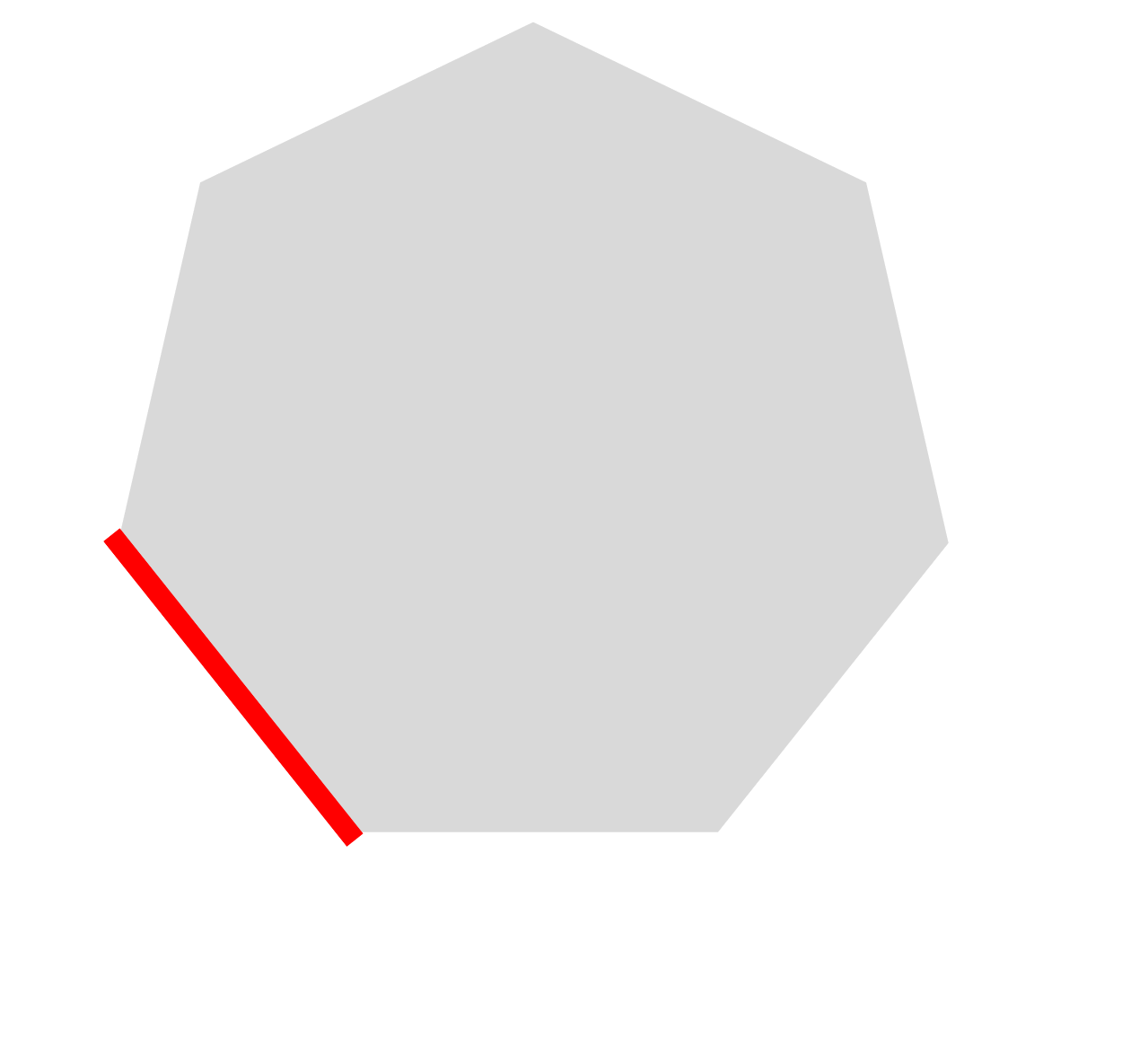}}&$n$&$1$\\
\hline
$0$&\adjustbox{raise=-0.25cm}{\includegraphics[scale=0.05]{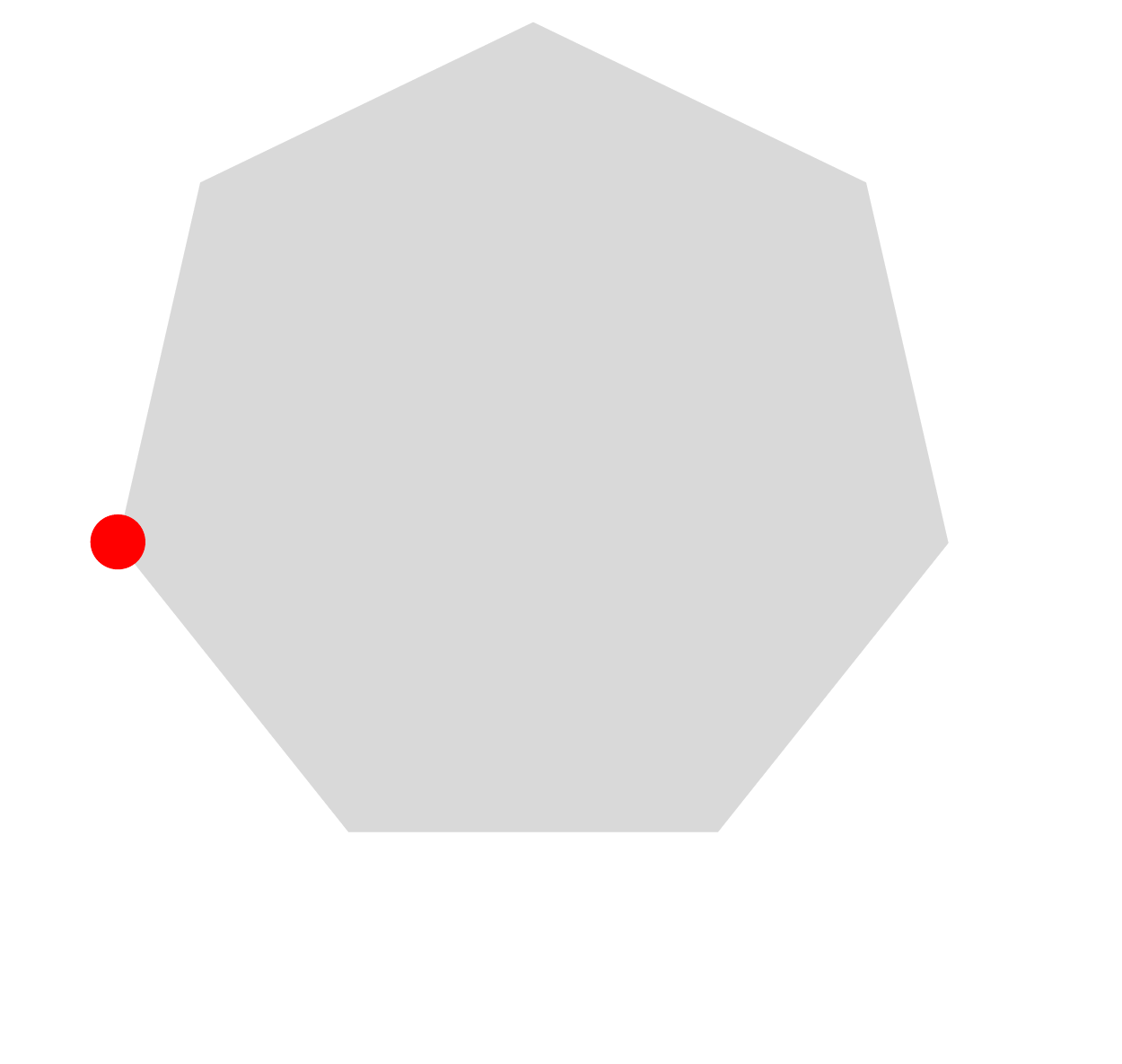}}&$n$&$0$\\
\hline
\end{tabular}

\bigskip
$ \chi_{n,1}=1-n+n=1$
\end{center}
\caption{All boundaries for the $k=1$ amplituhedron.}
\label{Tab:k1}
\end{table}
For $k=2$, there is again $n$ co-dimension one boundaries and also $n$ co-dimension two boundaries of the type $P^{(n)}_{top}\otimes P^{(n)}_{i}$. There is a second type of co-dimension two boundaries, $P^{(n)}_{i,i+1}\otimes P^{(n)}_{j,j+1}$, and since $i < j$ then there are $\binom{n}{2}$ of them. The number of $\dim_A=1$ boundaries $P^{(n)}_{i,i+1}\otimes P^{(n)}_{j}$ is $n(n-1)$ since $i\neq j,j+1$ and we need to remember to include the non-generic boundary $P^{(n)}_{i-1,i+1}\otimes P^{(n)}_{i}$. Finally, there are $\binom{n}{2}$ zero-dimensional boundaries $P^{(n)}_{i}\otimes P^{(n)}_{j}$, with $i< j$. All results are collected in Table \ref{Tab:k2}.
\begin{table}[t]
\begin{center}
\begin{tabular}{|c|c|c|c|}
\hline
$\dim_A$&boundary type&number of boundaries&$\dim_C$\\
\hline
$4$&\adjustbox{raise=-0.25cm}{\includegraphics[scale=0.05]{polyfull.pdf}\,\,\includegraphics[scale=0.05]{polyfull.pdf}}&$1$&$2n-4$\\
\hline
$3$&\adjustbox{raise=-0.25cm}{\includegraphics[scale=0.05]{polyfull.pdf}\,\,\includegraphics[scale=0.05]{polylineclear.pdf}}&$n$&$n-1$\\
\hline
$2$&\adjustbox{raise=-0.25cm}{\includegraphics[scale=0.05]{polyfull.pdf}\,\,\includegraphics[scale=0.05]{polypointclear.pdf}}&$n$&$n-2$\\
\hline
$2$&\adjustbox{raise=-0.25cm}{\includegraphics[scale=0.05]{polylineclear.pdf}\,\,\includegraphics[scale=0.05]{polylineclear.pdf}}&$\binom{n}{2}$&$2$\\
\hline
$1$&\adjustbox{raise=-0.25cm}{\includegraphics[scale=0.05]{polylineclear.pdf}\,\,\includegraphics[scale=0.05]{polypointclear.pdf}}&$n(n-1)$&$1$\\
\hline
$0$&\adjustbox{raise=-0.25cm}{\includegraphics[scale=0.05]{polypointclear.pdf}\,\,\includegraphics[scale=0.05]{polypointclear.pdf}}&$\binom{n}{2}$&$0$\\
\hline
\end{tabular}

\bigskip 
 $ \chi_{n,2}=1-n+n+\binom{n}{2}-n(n-1)+\binom{n}{2}=1$
\end{center}
\caption{All boundaries for the $k=2$ amplituhedron. }
\label{Tab:k2}
\end{table}
We also include $k=3$ results in Table \ref{Tab:k3}. Results for higher $k$ can be easily generated from our diagrammatics. Importantly, in all cases we studied we found that the Euler characteristic of $\mathcal{A}_{n,k}^{(2)}$ equals 1. We provide a general argument for this assertion in the following section.
\begin{table}[h!]
\begin{center}
\begin{tabular}{|c|c|c|c|}
\hline
$\dim_A$&boundary type&number of boundaries&$\dim_C$\\
\hline
$6$&\adjustbox{raise=-0.25cm}{\includegraphics[scale=0.05]{polyfull.pdf}\,\,\includegraphics[scale=0.05]{polyfull.pdf}\,\,\includegraphics[scale=0.05]{polyfull.pdf}}&$1$&$3n-9$\\
\hline
$5$&\adjustbox{raise=-0.25cm}{\includegraphics[scale=0.05]{polyfull.pdf}\,\,\includegraphics[scale=0.05]{polyfull.pdf}\,\,\includegraphics[scale=0.05]{polylineclear.pdf}}&$n$&$2n-5$\\
\hline
$4$&\adjustbox{raise=-0.25cm}{\includegraphics[scale=0.05]{polyfull.pdf}\,\,\includegraphics[scale=0.05]{polyfull.pdf}\,\,\includegraphics[scale=0.05]{polypointclear.pdf}}&$n$&$2n-6$\\
\hline
$4$&\adjustbox{raise=-0.25cm}{\includegraphics[scale=0.05]{polyfull.pdf}\,\,\includegraphics[scale=0.05]{polylineclear.pdf}\,\,\includegraphics[scale=0.05]{polylineclear.pdf}}&$\binom{n}{2}$&$n-1$\\
\hline
$3$&\adjustbox{raise=-0.25cm}{\includegraphics[scale=0.05]{polyfull.pdf}\,\,\includegraphics[scale=0.05]{polylineclear.pdf}\,\,\includegraphics[scale=0.05]{polypointclear.pdf}}&$n(n-1)$&$n-2$\\
\hline
$3$&\adjustbox{raise=-0.25cm}{\includegraphics[scale=0.05]{polylineclear.pdf}\,\,\includegraphics[scale=0.05]{polylineclear.pdf}\,\,\includegraphics[scale=0.05]{polylineclear.pdf}}&$\binom{n}{3}$&$3$\\
\hline
$2$&\adjustbox{raise=-0.25cm}{\includegraphics[scale=0.05]{polyfull.pdf}\,\,\includegraphics[scale=0.05]{polypointclear.pdf}\,\,\includegraphics[scale=0.05]{polypointclear.pdf}}&$\binom{n}{2}$&$n-3$\\
\hline
$2$&\adjustbox{raise=-0.25cm}{\includegraphics[scale=0.05]{polylineclear.pdf}\,\,\includegraphics[scale=0.05]{polylineclear.pdf}\,\,\includegraphics[scale=0.05]{polypointclear.pdf}}&$n\cdot \binom{n-1}{2}$&$2$\\
\hline
$1$&\adjustbox{raise=-0.25cm}{\includegraphics[scale=0.05]{polylineclear.pdf}\,\,\includegraphics[scale=0.05]{polypointclear.pdf}\,\,\includegraphics[scale=0.05]{polypointclear.pdf}}&$\binom{n}{2}\cdot (n-2)$&$1$\\
\hline
$0$&\adjustbox{raise=-0.25cm}{\includegraphics[scale=0.05]{polypointclear.pdf}\,\,\includegraphics[scale=0.05]{polypointclear.pdf}\,\,\includegraphics[scale=0.05]{polypointclear.pdf}}&$\binom{n}{3}$&$0$\\
\hline
\end{tabular}

\bigskip
 $ \chi_{n,3}=1-n+n+\binom{n}{2}-n(n-1)-\binom{n}{3}+\binom{n}{2}+n\cdot\binom{n}{2}-\binom{n}{2}\cdot (n-2)+\binom{n}{3}=1$
\end{center}
\caption{All boundaries for the $k=3$ amplituhedron. }
\label{Tab:k3}
\end{table}

\subsection{Generating Function}
To make a general claim about the number of amplituhedron boundaries and the amplituhedron Euler characteristic, it is instructive to introduce a generating function:
\begin{equation}
\mathcal{F}_{n,k}(x,y)=\sum_{B\in\mathcal{B}_{n,k}}  \,(-x)^{\text{codim}_AB}y^{\#_B\left[\includegraphics[scale=0.02]{polyfull.pdf}\right]}\,,
\end{equation}
where the power of $x$ counts the amplituhedron co-dimension of the boundary $B\in \mathcal{B}_{n,k}$, where $\text{codim}_A\, B=2k-\dim_A B$, and the power of $y$ counts how many solid $n$-gons there are in the graphical representation of $B$. By collecting data from the previous section we can write an explicit form of this function for $k=1,2,3$:
\begin{align}
\mathcal{F}_{n,1}(x,y)&=y-\binom{n}{1}\,x+\binom{n}{1}\,x^2=y-\binom{n}{1}\,x(1-x)\,,\\\nonumber
\mathcal{F}_{n,2}(x,y)&=y^2-\binom{n}{1}\,y\,x+\binom{n}{1}\,y\,x^2+\binom{n}{2}\, x^2-2\binom{n}{2}x^3+\binom{n}{2}x^4\\
&=y^2-\binom{n}{1}\,y\,x(1-x)+\binom{n}{2}\,x^2(1-x)^2\,,\\\nonumber
\mathcal{F}_{n,3}(x,y)&=y^3-\binom{n}{1}\,y^2\,x+\binom{n}{1}\,y^2\,x^2+\binom{n}{2}\,y\, x^2-2\binom{n}{2}\,y\,x^3-\binom{n}{3}\,x^3\\\nonumber
&+\binom{n}{2}\,y\,x^4+3\binom{n}{3}\,x^4-3\binom{n}{3}\,x^5+\binom{n}{3}\,x^6\\
&=y^3-\binom{n}{1}\,y^2\,x(1-x)+\binom{n}{2}\,y\,x^2(1-x)^2-\binom{n}{3}\,x^3(1-x)^3\,.
\end{align}
By studying more examples for higher $k$, one can find a general formula valid for any $k$:
\begin{equation}\label{gen.fun.gen}
\mathcal{F}_{n,k}(x,y)=\sum_{i=0}^{k}(-1)^{i}\binom{n}{i}\,y^{k-i}x^{i}(1-x)^{i}\,.
\end{equation}
The Euler characteristic can be extracted from the function $\mathcal{F}_{n,k}$ by evaluating it at $x= 1$ and $y= 1$. It is easy to notice that only the first term in the sum \eqref{gen.fun.gen} survives and we find for all $n$ and $k$ 
\begin{equation}
\chi_{n,k}=\mathcal{F}_{n,k}(1,1)=1\,.
\end{equation}

\subsection{Amplituhedron Boundary Operator} 
We finish this section by constructing an {\it amplituhedron boundary operator} $\partial_A$ with the property
\begin{equation}\label{partialsqr}
\partial_A\circ\partial_A=0\,.
\end{equation} 
We define the following action of $\partial_A$ on a single-site $n$-gon label:
\begin{align}
&\partial_A P^{(n)}_{top}=\sum_{i=1}^nP^{(n)}_{ii+1}\,,\\
&\partial_A P^{(n)}_{ii+1}=P^{(n)}_{i}-P^{(n)}_{i+1}\,,\\
&\partial_AP^{(n)}_{i}=0\,,
\end{align}
and extend it for any $k$ as
\begin{equation}\label{del.action.k}
\partial_A( \mathcal{P}_1\otimes\ldots\otimes \mathcal{P}_k) = \sum_{a=1}^k s_a \mathcal{P}_1\otimes\ldots\otimes \partial_{A}\mathcal{P}_a\otimes\ldots\otimes \mathcal{P}_k\,,
\end{equation}
where $\mathcal{P}_a$ is one of the labels $\left(P^{(n)}_{top},P^{(n)}_{i,i+1},P^{(n)}_{i}\right)$ and the sign $s_a$ is determined by treating $\partial_A$ and $P^{(n)}_{i,i+1}$ as Grassmann-odd symbols (and treating $P^{(n)}_{top}$ as Grassmann-even). To check that \eqref{partialsqr} holds true, it is enough to check the action of $\partial_A^2$ on a two-fold product $\mathcal{P}_1\otimes \mathcal{P}_2$. The only non-trivial calculation to be done is in one of two cases: $P^{(n)}_{top}\otimes P^{(n)}_{top}$ or $P^{(n)}_{top}\otimes P^{(n)}_{i,i+1}$. It is straightforward to check that:
\begin{align}
\partial_{A}\left(\partial_A\left( P^{(n)}_{top}\otimes P^{(n)}_{top}\right)\right)&=2\partial_A\left(\sum_{i=1}^n P^{(n)}_{top}\otimes P^{(n)}_{i,i+1}\right)\\
&\hspace{-1cm}=2\left(\sum_{i} P^{(n)}_{top}\otimes\left(P^{(n)}_{i}-P^{(n)}_{i+1}\right)-\sum_{i,j=1}^nP^{(n)}_{j,j+1}\otimes P^{(n)}_{i,i+1}\right)=0\,,
\end{align}
where the first term vanishes because $\sum\limits_{i}\left(P^{(n)}_{i}-P^{(n)}_{i+1}\right)=0$ and in the second we used the fact that $P^{(n)}_{i,i+1}$ are Grassmann-odd symbols and therefore $P^{(n)}_{i,i+1}\otimes P^{(n)}_{j,j+1}=-P^{(n)}_{j,j+1}\otimes P^{(n)}_{i,i+1}$. In the second case we get:
\begin{align}
&\partial_{A}\left(\partial_A\left( P^{(n)}_{top}\otimes P^{(n)}_{i,i+1}\right)\right)=\partial_A\left(\sum_{j=1}^n P^{(n)}_{j,j+1}\otimes P^{(n)}_{i,i+1}+ P^{(n)}_{top}\otimes \left(P^{(n)}_{i}-P^{(n)}_{i+1}\right)\right)\\\nonumber
&=\sum_{j=1}^n \left(P^{(n)}_{j}-P^{(n)}_{j+1}\right)\otimes P^{(n)}_{i,i+1}-\sum_{j=1}^n P^{(n)}_{j,j+1}\otimes \left(P^{(n)}_{i}-P^{(n)}_{i+1}\right)+\sum_{j=1}^n P^{(n)}_{j,j+1}\otimes \left(P^{(n)}_{i}-P^{(n)}_{i+1}\right)\\
&=0\,,
\end{align}
where the minus sign in the second line comes from $s_a$ in the definition \eqref{del.action.k}. The remaining cases trivially vanish, which proves the formula \eqref{partialsqr}.

We can use the boundary operator $\partial_A$ to generate all co-dimension one boundaries of $\mathcal{A}_{n,k}^{(2)}$ by acting on $P^{(n)}_{top}\otimes\ldots\otimes P_{top}^{(n)}$. We can proceed recursively and generate all boundaries with the amplituhedron dimension equal $d$ by acting with $\partial_A$ on all boundaries of dimension $d+1$, which we found in the previous recursive step. In order to get the complete agreement with our classification of boundaries, we need to additionally assume that
\begin{align}
\left(P^{(n)}_{i,j}-P^{(n)}_{i,l}+P^{(n)}_{j,l}\right) \otimes P^{(n)}_j=0\qquad \text{for }i<j<l=1,\ldots,n \,.
\end{align}
This allows us to find all boundaries of $\mathcal{A}_{n,k}^{(2)}$, including the non-generic ones of the form $P^{(n)}_{i-1,i+1}\otimes P^{(n)}_i$.
It remains an interesting open problem to find a simpler description of the boundary operator $\partial_A$, one which is  more similar to the positive Grassmannian boundary operator that acts as a transposition on the permutation associated to a cell.

\section{Conclusions and Outlook}
In this paper we studied the combinatorial structure of amplituhedron boundaries. We provided a diagrammatic description which classify all boundaries of $\mathcal{A}_{n,k}^{(2)}$, of all dimensions, for all $n$ and $k$.  This allows us to describe the amplituhedron boundary poset,  which we check to be Eulerian. We also found that the Euler characteristics for all amplituhedra equals one, which is a necessary condition for the amplituhedron $\mathcal{A}_{n,k}^{(2)}$ to be homeomorphic to a $(2\cdot k)$-dimensional closed ball.

There are obvious directions in which we can extend our results. It would be of great interest for physics community to analyze the case $m=4$, which is relevant for tree-level scattering amplitudes in planar $\mathcal{N}=4$ SYM. The boundaries of the amplituhedron $\mathcal{A}_{n,k}^{(4)}$ have not been classified yet and its topological properties are still unknown. Moreover, a similar analysis can be done for the recently introduced momentum amplituhedron \cite{Damgaard:2019ztj}. There also exists a further generalization of the amplituhedron to the so-called loop amplituhedron, which is a (still mostly unstudied)  positive geometry relevant for perturbation theory of planar $\mathcal{N}=4$ SYM, see \cite{Arkani-Hamed:2013kca,Franco:2014csa,Bai:2015qoa,Galloni:2016iuj,Arkani-Hamed:2018rsk,Langer:2019iuo} for known results. In this case, there are indications that the loop amplituhedron is not a ball anymore. However, it would be worth studying its complete boundary structure in a systematic way, which would provide us with a better understanding of integrands relevant for scattering amplitudes, as well as promise new results in combinatorics. 

\acknowledgments

We would like to thank Livia Ferro for useful discussions.
\appendix
\section{Properties of Boundaries from Graphical Notation}
\label{App:labels}

\noindent {\bf Amplituhedron dimension:} 
Given the label \eqref{general_label} for a boundary $B_\sigma\in\mathcal{B}_{n,k}$, the amplituhedron dimension of $\sigma$ can be calculated as:
\begin{equation}
\dim_A(\sigma)=2\cdot \#\left[P^{(n)}_{top}\right]+1\cdot \#\left[P^{(n)}_{ii+1}\right]+0\cdot \#\left[P^{(n)}_{i}\right]=2t-l\,.
\end{equation}

\noindent {\bf Grassmannian dimension:} 
Given the label \eqref{general_label} for a boundary $B_\sigma\in\mathcal{B}_{n,k}$, the amplituhedron dimension of $\sigma$ can be calculated as:
\begin{equation}
\dim_C(\sigma)=(n-k)t+l\,.
\end{equation}

\noindent {\bf Permutation:} Given the label \eqref{general_label} for a boundary $B_\sigma\in\mathcal{B}_{n,k}$,  we can find a permutation of the corresponding positroid cell $\sigma$ in the positive Grassmannian $G_{+}(k,n)$, which is mapped to $B_\sigma$ through the map $\Phi_Z$. To do that, let us construct a $k\times n$ matrix with entries $\{x_{ab}\}_{a=1,\ldots,k,b=1,\ldots,n}$. We set generic values to entries:
\begin{itemize}
\item for each $P^{(n)}_{top_a}$: $x_{ab}\neq 0$ for $a=1,\ldots,t$, $b=1,\ldots,n$,
\item for each $P^{(n)}_{i_ai_a+1}$: $x_{ab}\neq 0$ for $a=t+1,\ldots,t+l$, $b=i_a$ and $b=i_a+1$,
\item for each $P^{(n)}_{i_a}$: $x_{ab}\neq 0$ for $a=t+l+1,\ldots,k$, $b=i_a$.
\end{itemize}
and set all remaining entries $x_{ab}$ to zero. Then one can use the function {\tt matToPerm} from the Mathematica positroid package \cite{Bourjaily:2012gy} to find the permutation for the  positroid cell $\sigma$.

\bibliographystyle{nb}

\bibliography{ball}
\newpage
\newgeometry{margin=0pt,left=4cm,top=3cm}
\begin{figure}[ht!]
\begin{center}
\includegraphics[scale=0.55]{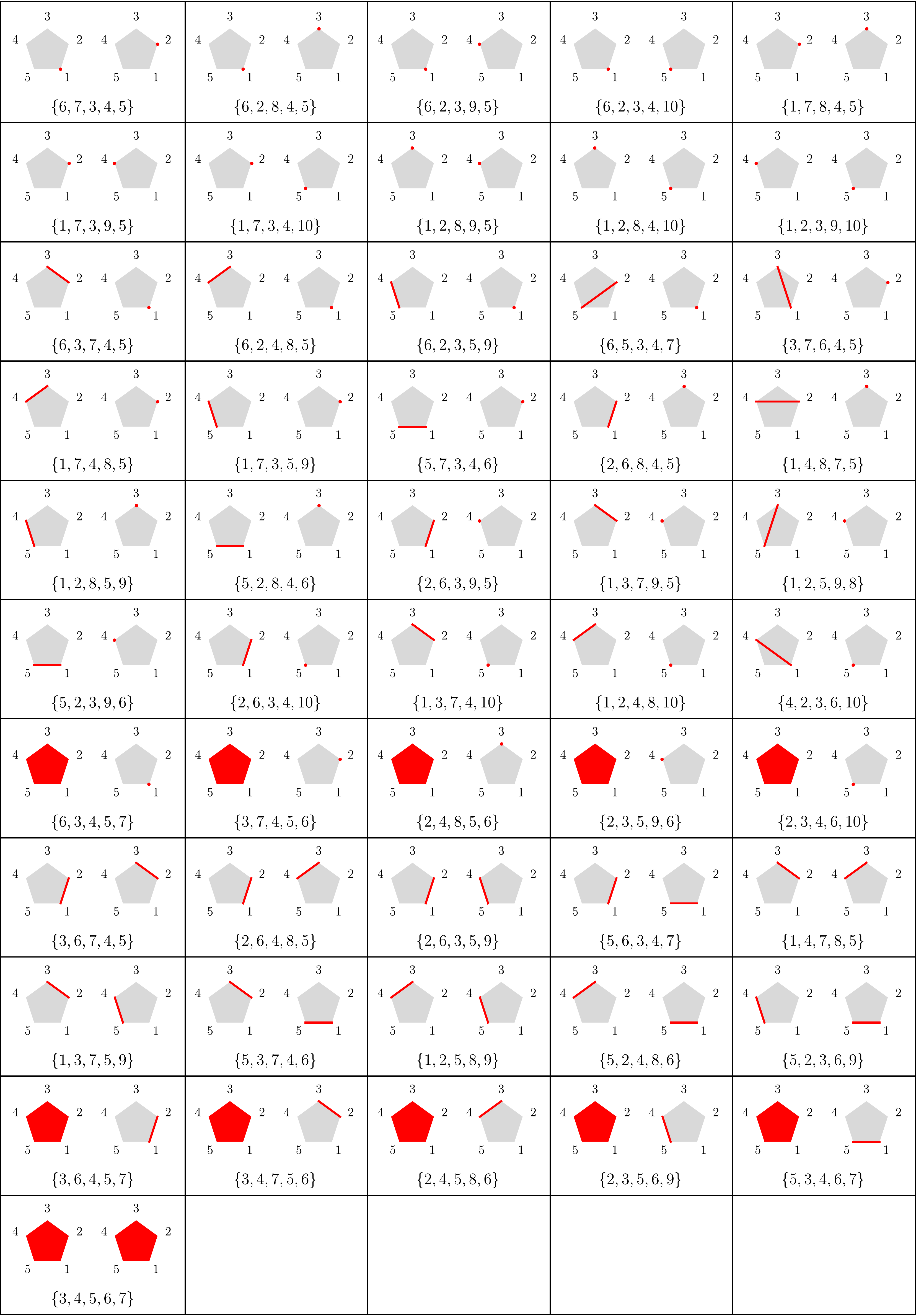}
\end{center}
\caption{All labels for boundaries of  $\mathcal{A}_{5,2}^{(2)}$ with the permutation of corresponding positroid cells.}
\label{Fig:n5k2}
\end{figure}

\newpage
\newgeometry{margin=0pt,left=4cm,top=3cm}
\begin{figure}
\begin{center}
\includegraphics[scale=0.8,angle=90]{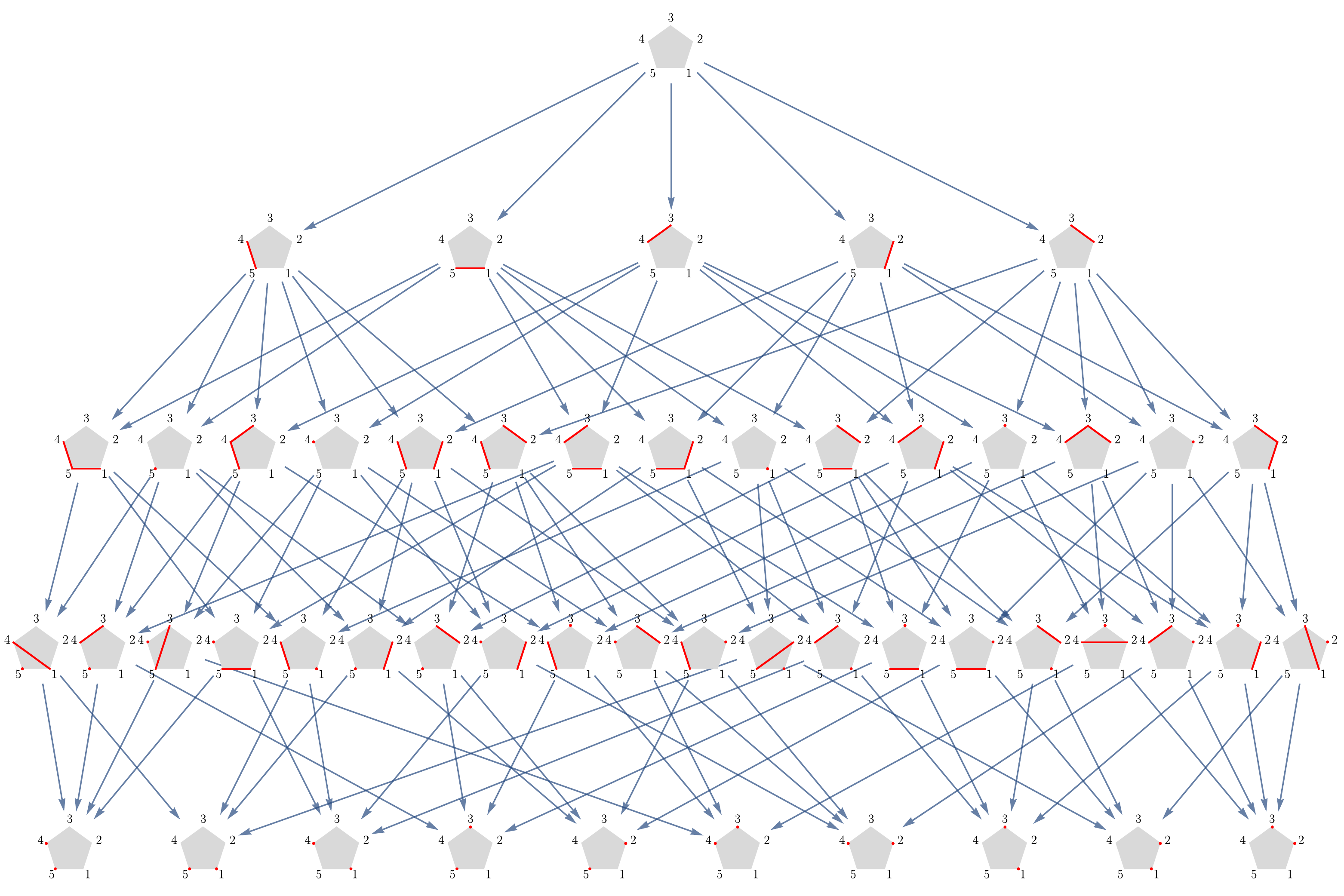}
\end{center}
\caption{Poset of boundaries for $\mathcal{A}_{5,2}^{(2)}$, using the reduced diagrammatical notation.}
\label{Fig:posetn5k2}
\end{figure}

\end{document}